\documentclass[%
 reprint,
 longbibliography,
 amsmath,amssymb,
 aps,
]{revtex4-1}

\usepackage{graphicx}
\usepackage{dcolumn}
\usepackage{bm}

\usepackage[utf8]{inputenc}
\usepackage[english]{babel}

\newtheorem{conjecture}{Conjecture}

\begin{document}


\title{Quantum Mechanics as Naturalized Time}


\author{Zachary D. Walton}
 \email{zachary.walton@gmail.com}

\author{Bernard S. McNamara}

\author{Tommaso Toffoli}
\email{tt@bu.edu}


\date{\today}

\begin{abstract}
We bring together two topics (quantum mechanics and time passage) with the goal of clarifying questions about each.  Specifically, we claim that the formalism of quantum mechanics provides an answer to the question: ``What is time passage?''. 
\end{abstract}

\pacs{Valid PACS appear here}
\maketitle


\section{\label{sec:level1}Introduction}

What is time passage?  Why is quantum mechanics a successful theory of the world?  We claim that by addressing the first question, we gain insight into the second.  In particular, we propose a mathematical correlate to time passage, and show that this construction, along with uncontroversial operational assumptions, place constraints on the kind of physics an observer might see.  Finally, we provide support for the conjecture that these constraints make quantum mechanics a forced move.

Time's direction and time passage are two related but distinct issues in the philosophy of time.  The novel construction we propose as a path to quantum mechanics relates to time passage.  However, since time passage must be defined relative to time's direction, time's direction logically precedes time passage.  In practical terms, we imagine looking at a pile of unordered photographs taken of a laboratory during an experiment.  We further imagine that the experimenter has a glass of ice water on the table throughout the experiment.  The progress of ice cubes melting in the glass provides a reliable proxy for how the photographs should be ordered, past to present.  The question of time passage relates to how we understand the experimenter to be the same person in each photograph, beyond spatiotemporal proximity in a geometrized space-time.    

Our argument hinges on the question of what counts as an explanation of the common sense notion of time passage. There is a tension in philosophy of physics between realists, who seek in physics an observer-independent story of what is going on, and instrumentalists, who view physics as nothing more than an exercise in organizing the records of observers' experiences. Even the most staunch realist will concede that a physical theory must have implications for our experience, expressed in statements such as 

\begin{multline}
\text{``I\textsubscript{before} prepared the system in state $S$;}\\
\text{then I\textsubscript{after} performed measurement $M$,}\\
\text{which resulted in outcome $O$.'' }    
\label{experimental-log}
\end{multline}

While it is not customary to use distinct symbols for I\textsubscript{before} and I\textsubscript{after}, we do so here in order to probe the role (if any) played by time passage.  To what degree must a physical theory explain the relationship between I\textsubscript{before} and I\textsubscript{after}?  The block universe (``timeless'') picture of the world asserts that I\textsubscript{before} and I\textsubscript{after} are distinct slices of the assemblage of world lines traced by the particles in the observer's body.  In this view, time passage does not have a correlate, and if we are surprised by this state of affairs, it is because we were confused about passage to begin with (``passage as illusion'').

Does time passage need a correlate in physical theory?  To aid our intuition regarding what an answer to this question might look like, we consider the example of color.  Color has a correlate in physical theory (i.e., the frequency of electromagnetic waves).  Before Maxwell solidified the connection between light and electromagnetism, it was possible to believe that color is fundamental (i.e., not reducible to any other aspect of our experience).  However, in this scenario, it would have been a mistake to assert that ``color is an illusion''.  

Maxwell's equations explain one aspect of human experience (color) in terms of another (electromagnetism).  That is, color was {\em naturalized} by Maxwell's equations. This is the normal sequence of events in the scientific process.  An aspect of human experience (e.g., color) has the status of a primitive of reality, until a connection to a distinct aspect of human experience (e.g., electromagnetism) is discovered.  The explanatory power of this connection allows deeper understanding of both sides of the connection, and puts to rest the idea that the explained phenomenon was an illusion.

Maxwell brought together areas of human experience that a priori seem very different: color vision and electromagnetism.  Evidence of the distance he bridged in conceptual space is provided by the persistent suspicion that there are aspects of color vision that go beyond physics (c.f. qualia and the Mary's Room thought experiment~\cite{dennett}).  If, as we claim, time passage is like color, then time passage should have a mathematical correlate.  For color, the mathematical correlate is Maxwell's equations.  We shall argue that the mathematical correlate for time passage leads directly to the structure of quantum mechanics.  If we succeed, the payoff is two-fold: insight into the nature of time, and an answer to Wheeler's question ``Why the quantum?''\cite{wheeler}.

\section{Other work linking time passage and quantum mechanics}

We will refer to several papers in which connections (either direct or via analogy) are identified between quantum mechanics and time passage.  However, an important reference for our approach is one that focuses only on time.  Tim Maudlin has championed the minority view that it is a mistake to accept physics without time passage~\cite{maudlin}.  Rather, he argues, time passage should be regarded as fundamental.  His analysis is important for our effort because it represents a rare voice dissenting from the prevailing view of time as illusion. We and Maudlin agree that time passage must appear in our theory of the world.  However, while Maudlin takes time passage to be fundamental, we assert that time passage can be explained in terms of other aspects of our experience.

Simon Saunders presented an extended analogy between controversies in quantum mechanics and philosophy of time~\cite{saunders}.  His motivation was to defend the Everett interpretation from the following criticism: if no element of the superposition is selected to be more real than any other during a measurement event, there is no way to make sense of probabilities.  He points out that the arguments criticizing Everett's lack of a mechanism for identifying one outcome as actual are no stronger than the corresponding critique of the block universe, in which there is no mechanism for picking out one time as Now.  Thus, it is inconsistent to take on board the block universe while rejecting Everett on these grounds. Saunders argues that the block universe and Everett rise and fall together, with regard to the question of whether something is missing.  While his position is that they rise together, ours is that they do, in fact, miss something.  Nonetheless, his argument is relevant to our approach because it leverages anxiety about the extravagance of many worlds to sharpen the focus on the problem of time passage, which, having remained inscrutable for thousands of years, is often regarded as insoluble or not even a problem.

N. David Mermin also identifies a kinship between the interpretation of quantum mechanics and the philosophy of time~\cite{mermin}.  In particular, he argues that progress made by Chris Fuchs and R{\"u}diger Schack in the ``QBism'' project~\cite{fuchs} also serves to calm anxiety about the absence of the Now in the block universe.  This argument follows a similar architecture to that of Saunders (collapse-less quantum mechanics and the Now-less block universe rise together), except it leverages the arguments in the other direction.  In other words, our potential comfort with progress in the quantum realm should soothe our anxieties about the block universe.  

Saunders and Mermin were concerned with analogies between arguments in quantum mechanics and the philosophy of time, not a direct link.  In contrast, William Wootters directly implicates time in the origin of quantum mechanics~\cite{wootters}.  He argues that the raw material of quantum mechanics is not complex amplitudes, but rather measurement results.  While complex amplitudes are described as undergoing time-symmetric evolution, measurement results are laden with time asymmetry, as evidenced by the before/after distinction implicit in statement~(\ref{experimental-log}).  Wootters stops short of proposing a mathematical representation of this time asymmetry from which quantum mechanics arises.  Nonetheless, he does speculate that the identification of this time asymmetry in the resulting physical picture will not have the fundamental objective status that we might expect from a new piece of physics.  Rather, the choice involved in partitioning the world into subject and object will introduce a fundamental flexibility that thwarts ``the goal of mirroring perfectly in one's theory an independently existing reality''~\cite{wootters}.  Notwithstanding the threatened loss of objective reality, the realist will be heartened that a coherent conception of reality can survive the program suggested by Wootters and developed here.  In fact, we claim that by clarifying the notions of time passage and the origins of quantum mechanics, we may place the necessary concepts ``system'' and ``observer'' on surer footing than they rest with either the traditional realist or the instrumentalist positions. 

\section{A link between quantum mechanics and personal identity}

The central work of this paper is our attempt to derive quantum mechanics from a proposed mathematical correlate to time passage.  To motivate this association, we argue that the formalism of quantum mechanics contains the right kind of structure to enable us to make sense of time passage.  Consider the following version of the Wigner's friend thought experiment, in which Alice is performing experiments on a qubit, and Bob has coherent control over the joint system of Alice and the qubit~\cite{wigner}.  Alice has prepared a pure qubit, characterized by a pair of real numbers (e.g., the latitude and longitude on the Bloch sphere).  Now she performs a measurement in the basis corresponding to the north-south axis.  What is the state of the qubit?  It depends on who you ask.  For Alice, the qubit is measured to be either due north or due south.  For Bob, the entire Alice-qubit system remains a pure state in which the qubit is entangled with Alice.

The formalism of quantum mechanics encodes a nontrivial link between personal identity and state determination.  This link distinguishes quantum from classical physics.  In classical physics, the world may look different to observers with different knowledge, but the formalism allows for these distinct states of knowledge to refer to the same underlying reality.  In quantum mechanics, the two observers have different views of the world even before the notion of incomplete knowledge is introduced.  When asking questions of classical mechanics, we must first answer ``what do you know?'', while in asking questions of quantum mechanics, we must first answer ``who are you?''.  Put another way, time passage (i.e., how the two I's in statement (\ref{experimental-log}) are related) involves concepts that touch on a feature that distinguish quantum mechanics from classical mechanics.  Different people may have distinct views of reality, even accounting for differences in knowledge.  

\section{Extending the link to time passage}

Staying with the Wigner's friend example, we observe that when Alice measures her qubit along the north-south axis, the latitude of the state of the qubit influences her measurement result, but the longitude does not.  But just because this longitude degree of freedom is lost to Alice's results, is it in fact lost to everyone?  Clearly not.  Bob can, at any point, reverse Alice's measurement and measure the qubit in a basis other than that corresponding to the north-south axis.  In performing this reversal, Bob will necessarily remove any evidence of Alice's measurement result (``north'' or ``south''), even if that information had been amplified to the point at which it is represented in Alice's brain state.  This erasure is a fundamental feature of quantum mechanics, and it appears in many contexts, such as the trade-off between which-path information and the quality of the interference pattern in the double-slit experiment.  The conclusion relevant for our purposes is that Alice may perform her measurement and assert that the longitude information is lost without fear of being proven wrong, since Bob's procedure for recovering the longitude information will necessarily remove any record of her claim.  Restating in the notation of statement (\ref{experimental-log}), Alice\textsubscript{after} may assert that the longitude information is lost, because the procedure for revealing it will effectively erase any record of her existence.  We label this property of quantum mechanics the ``collapse defense principle (CDP),'' since it guarantees that Alice may confidently assert that the wavefunction has collapsed (for her) without fear of being proven wrong.

We view the CDP as akin to Einstein's equivalence principle (i.e., the connection between acceleration and gravity).  A priori, it seems like a coincidence that accelerating in an elevator in empty space is indistinguishable from the effects of a uniform gravitational field.  Einstein recast this indistinguishability as a foundational principle (i.e., they are indistinguishable, because at a deep level they are the same).  Instead of simply observing that quantum mechanics implies that longitude information is lost to Alice\textsubscript{after}, we propose {\em defining} Alice\textsubscript{after} as Alice\textsubscript{before} with the condition that longitude information is lost.  More generally, observe that in choosing a basis for a von Neumann measurement of a quantum system, we partition the degrees of freedom of a pure quantum state into those that are sampled and those that are lost (the latitude and the longitude, respectively, in Alice's case). We propose defining the post-measurement observer as that person for whom the non-sampled degrees of freedom are lost (``you are what you lose''). 


We began by arguing that we need a correlate to time passage, just as we needed a correlate to color before Maxwell.  We argued that quantum mechanics has the right kind of structure to identify a connection to time passage (``what you see depends on who you are'').  Then we recast a specific property of quantum mechanics (i.e., certain degrees of freedom are lost to Alice during measurement) as a definition of the relationship between Alice\textsubscript{before} and Alice\textsubscript{after} (i.e., Alice\textsubscript{after} is the version of Alice\textsubscript{before} for whom those degrees of freedom are lost).  In describing the relationship between Alice\textsubscript{before} and Alice\textsubscript{after}, we are providing a candidate correlate for time passage.  

It is audacious to suggest that a problem contemplated by the ancients (``what is time passage?'') is answered by the formalism of quantum mechanics (``you are what you lose'').  Two points may assuage the shock of this proposal.  First, before Maxwell, it would have been similarly audacious to propose that color experience was reducible to a particular kind of wave behavior of electric and magnetic fields.  Second, there is a mathematical argument that makes the connection between time passage and quantum mechanics more plausible.  We will argue that a particularly natural model of time passage makes quantum mechanics a forced move.  If it can be demonstrated that this is a forced move, it will add credibility to the association of time passage with quantum mechanics, just as the validity of Maxwell's equations was bolstered by the fact the predicted speed of electromagnetic waves matched the measured speed of light.  

Even lacking the demonstration that quantum mechanics is a forced move, the association of time passage with the formalism of quantum mechanics is significant for three reasons.  First, if we grant that time passage needs a correlate, the dearth of candidates makes any proposition worth considering.   Second, we have avoided the trap of begging the question, which is the mistake one makes when imagining that a laser pointer moving its focus along a world line provides an explanation of time passage.  That is, we propose explaining time passage in terms of a distinct aspect of human experience (i.e., the mathematics of quantum mechanics).  Third, the specific link we have made (``you are what you lose'') satisfies the base requirement that it agrees with quantum mechanics in its depiction of the relationship between Alice\textsubscript{before} and Alice\textsubscript{after}.  The CDP ensures operational equivalence between the conventional quantum mechanical description of the post-measurement observer Alice\textsubscript{after} and our identification of Alice\textsubscript{after} as a version of Alice\textsubscript{before} for whom the information in certain degrees of freedom is lost. 

We claim that quantum mechanics provides an answer to the question ``what is time passage?''.  Historically, quantum mechanics was developed to address specific experimental results (black body radiation, atomic spectra, etc.).  However, if our claimed link between quantum mechanics and time passage is valid, one could imagine the mathematical structure of quantum mechanics being derived before these experimental effects were studied, using only concepts related to time passage.  This is our task for the remainder of the paper.

\section{Correlated selection}

What should a naturalized theory of time passage look like?  It should involve an observer and a system external to the observer, and these two entities should interact.  Thus, we ground our model of passage in the framework of measurement theory.  We consider the classical measurement story first, and then explore how it may be modified to include a naturalized conception of time passage.

Consider the familiar case of classical measurement.  We begin by focusing on dichotomous variables.  There is a physical system and an observer's memory, each of which can take on one of two states (``1'' or ``0'').  Measurement occurs when a specific dynamics is achieved, such that the state of the observer's memory is retained if the system is in state ``0'', and flipped if the system is in state ``1''.  That is,

\begin{equation} 
\begin{split}
(0_s,0_o) & \rightarrow (0_s,0_o) \\
(1_s,0_o) & \rightarrow (1_s,1_o) \\
(0_s,1_o) & \rightarrow (0_s,1_o) \\
(1_s,1_o) & \rightarrow (1_s,0_o) \\
\end{split}
\label{eq1}
\end{equation}
where the subscripts $s$ and $o$ refer to the system and the observer, respectively.  When the observer prepares the memory in state ``0'' (at some thermodynamic cost), this dynamics effectively copies the state of the system into the memory.  We see this by focusing only on the first two lines of (\ref{eq1}).  This model of measurement achieves the intuitively necessary effect of establishing correlation between the system and memory.  There are two parts of this model of classical measurement: the underlying dynamics embodied by (\ref{eq1}) and the initial probability distribution over the state space.  In order to achieve a natural embodiment of time passage, we must remove reference to ``dynamics,'' as this is the concept we are trying to explain.

Instead of dynamics, we propose {\em correlated selection}.  We imagine the initial state is one in which the system and the observer's memory are uncorrelated random variables (e.g., two unfair coins with distinct probabilities of coming up heads).  The final state is one with both variables having the same state, and we achieve this by removing the cases in which they disagree.  That is, we imagine repeatedly flipping two unfair coins (with probabilities of heads $p_s$ and $p_o$ for the system coin and the observer coin, respectively), crossing out occasions when the two coins have different results $[\mbox{i.e.}, (1_s,0_o)$ and $(0_s,1_o)]$, and measuring the relative frequencies of the two cases left over $[\mbox{i.e.}, (1_s,1_o)$ and $(0_s,0_o)]$.  This construction maps the two-dimensional probability space described by
\begin{equation} 
\begin{split}
(0_s,0_o) & \mbox{ with probability } \bar{p}_s\bar{p}_o \\
(1_s,0_o) & \mbox{ with probability } p_s\bar{p}_o \\
(0_s,1_o) & \mbox{ with probability } \bar{p}_sp_o \\
(1_s,1_o) & \mbox{ with probability } p_sp_o \\
\end{split}
\end{equation}
onto the one-dimensional probability space described by
\begin{equation} 
\begin{split}
(0_s,0_o) & \mbox{ with probability } \bar{p}_s\bar{p}_o/(p_sp_o + \bar{p}_s\bar{p}_o) \\
(1_s,0_o) & \mbox{ with probability } 0 \\
(0_s,1_o) & \mbox{ with probability } 0 \\
(1_s,1_o) & \mbox{ with probability } p_sp_o/(p_sp_o + \bar{p}_s\bar{p}_o) \\
\end{split}
\end{equation}
where $\bar{p}\equiv 1-p$.  The denominator in the non-zero terms serves to normalize the distribution.  For convenience, we introduce the function notation $\mbox{CS}(p_s,p_o)\equiv p_sp_o/(p_sp_o + \bar{p}_s\bar{p}_o)$, where $\mbox{CS}$ is meant to connote correlated selection.

Two observations are in order.  First, this model of time passage satisfies the base requirement that the before/after transition is characterized by a transition from uncorrelated system and observer to correlated system and observer.  Second, the dimension of the state spaces involved in correlated selection is the same as that in the collapse of the wavefunction.  Specifically, performing correlated selection on two unfair coins involves mapping a two-dimensional space onto a one-dimensional space, just as choosing a measurement basis for a pure qubit maps a point on a sphere to a point on a one-dimensional probability space. This result generalizes for N-state systems (i.e., correlated selection entails $2(N-1)$ degrees of freedom for the before state, just as a pure quantum state requires $2(N-1)$ degrees of freedom).  This agreement of state space size is a necessary (but not sufficient) condition for our derivation of quantum mechanics from time passage to go through.

\begin{figure}
\includegraphics[width=0.5\textwidth]{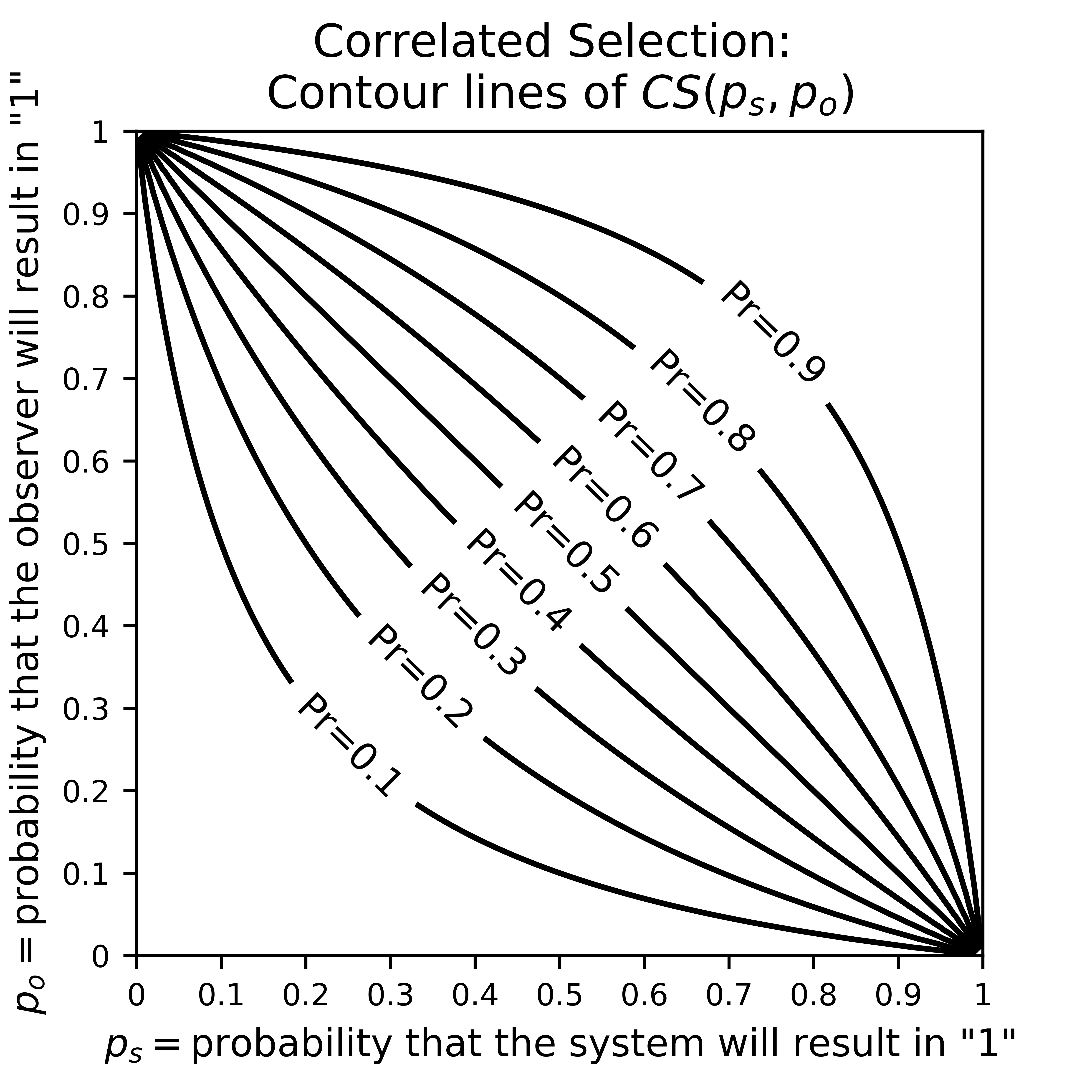}
\caption{The mechanism for time passage presented in this paper entails postselecting uncorrelated random variables corresponding to observer (x-axis) and system (y-axis), such that cases in which these two variables agree are retained.  This process leads to an outcome random variable which is ``1'' with probability $\mbox{CS}(p_s,p_o)$, illustrated above with contour lines.  The mechanism for time passage is completed by introducing a distance function on this unit square, as described in the text.}
\label{wootters}
\end{figure}

Correlated selection in the case $N=2$ amounts to a map from the unit square to the unit interval, and we can visualize the map with the contour plot in Fig.~\ref{wootters}. Notice that each contour line forms a connected one-dimensional subspace.  Our full model of time passage suggests that this unit square maps naturally onto a sphere, and that the contour lines in Fig.\ref{wootters} can be thought of as lines of latitude (i.e., lines of constant distance from the poles).  The fact that our model of time passage leads to a two-state system being described with a spherical state space (as opposed to toroidal, for example) is another necessary but not sufficient property for the purposes of linking to the Bloch sphere of quantum mechanics.  

We pause in our derivation of quantum mechanics from time passage to address a natural objection.  We are claiming that preparing a pure quantum state is in some sense equivalent to preparing a joint probability distribution over the random variables associated with the system and observer.  One might make the following objection: preparing a quantum state does not {\em feel} like preparing a joint probability distribution over the system and observer.  We reply that anytime a concept is explained, it is common to feel that something is lost and something is unwelcome.  For example, in the case of color, one is tempted to ask: ``But what is the actual color of the photon, putting aside its interactions with physical matter?''  Similarly, the problems brought along by Maxwell's equations (``What medium supports the waves?''), can make one doubt their utility if we lose sight of the explanatory work they do.  Likewise, asking for a dynamical underpinning of correlated selection (``Where did these excluded non-correlated cases go?'') or objecting to the mathematics proposed as the correlate for time passage (``Why correlated selection?'') are natural questions but should be put aside if we judge that the desired explanatory work has been accomplished.

\section{Preparation and Measurement}

Our model of time passage is not yet finished.  To see why, we observe that if the ``before'' state is characterized by two degrees of freedom (a point on the unit square) and the ``after'' state is characterized by one degree of freedom (a point on the unit interval), then two distinct ``before'' states that map to the same ``after'' state will be experimentally indistinguishable.  For this mechanism to match the experience of the observer, the two dimensions of the ``before'' state must be experimentally accessible.

Here we make our first of two appeals to the notion of reality.  For it to make sense for the observer to refer to an object, the observer must incorporate the notion of state preparation and state measurement.  An experimental apparatus typically has a number of controllable degrees of freedom, along with an output readable by the observer.  Without a distinction between state preparation and state measurement, the experimental process would amount to an unstructured lookup table.  Physics begins when the observer divides the controllable degrees of freedom into those used for preparing a state of the system and those used to specify the measurement.  In classical physics, this issue is barely worth pointing out, since the physical theory unambiguously identifies the reality underlying the observer's interaction with the system.  In quantum mechanics, the preparation/measurement issue is complicated by the lingering controversy about how the formalism is broken down into ontology and epistemology.  For example, in a quantum optics experiment, a single optical element may be considered part of state preparation or state measurement, with no change in the experimental results.  Notwithstanding this complication in the quantum case, a distinction between preparation and measurement is still necessary in any physics. 

Turning to our model of time passage, what mathematical construction could play the role of distinguishing between state preparation and measurement?  Our model associates the observer's description of the state of affairs before measurement with a point on the unit square.  State preparation and state measurement are closely allied, as each relates to the observer's view of the world.  State preparation conveys what the observer knows about the world, and state measurement conveys what the observer is asking about the world.  Thus, it is natural to mirror this duality in our model. 

Correlated selection is a mechanism for mapping a point on the unit square to a point on the unit interval (characterizing a probability distribution on the binary outcome variable).  To incorporate the notion of state preparation and state measurement, we propose extending this mechanism to entail a mapping between a {\em pair} of points on the unit square to a point on the unit interval.   We achieve this mechanism in two steps.  First, we geometrize the unit square by introducing a distance function $\mbox{D}$. Second, we use a probability function which maps distance to probability ($\mbox{Pr}_\textsubscript{D}$).  Specifically, we propose a distance function $\mbox{D}$ on the unit square such that points on the same contour line in Fig.~\ref{wootters} are equidistant from the point (1,1).  Defining $ \mbox{D}$ in this way constrains the probability function $\mbox{Pr}_\textsubscript{D}$.  For example, by traveling on the straight line between points (1,1) and (0,0), we may use $\mbox{D}$ and $\mbox{CS}(p_s,p_o)$ to generate probability function $\mbox{Pr}_\textsubscript{D}$.  

This step completes the specification of our proposed mathematical correlate of time passage.  In summary, we propose that time passage {\em is} correlated selection in the context of a geometrized joint probability space over observer and system.  We next illustrate this construction with a specific distance function and then argue for the forced move to quantum mechanics.

\section{Choosing a distance function}\label{distfunc}
Our model of time passage requires that a distance function be defined on the unit square such that the points along a given contour line in Fig.~\ref{wootters} are equidistant from the point (1,1).  Here we provide an example of such a distance function.  We then draw a connection between this example and quantum mechanics.  Finally, we argue that a necessary constraint on any such distance function makes this connection a forced move.

\begin{figure}
\includegraphics[width=0.5\textwidth]{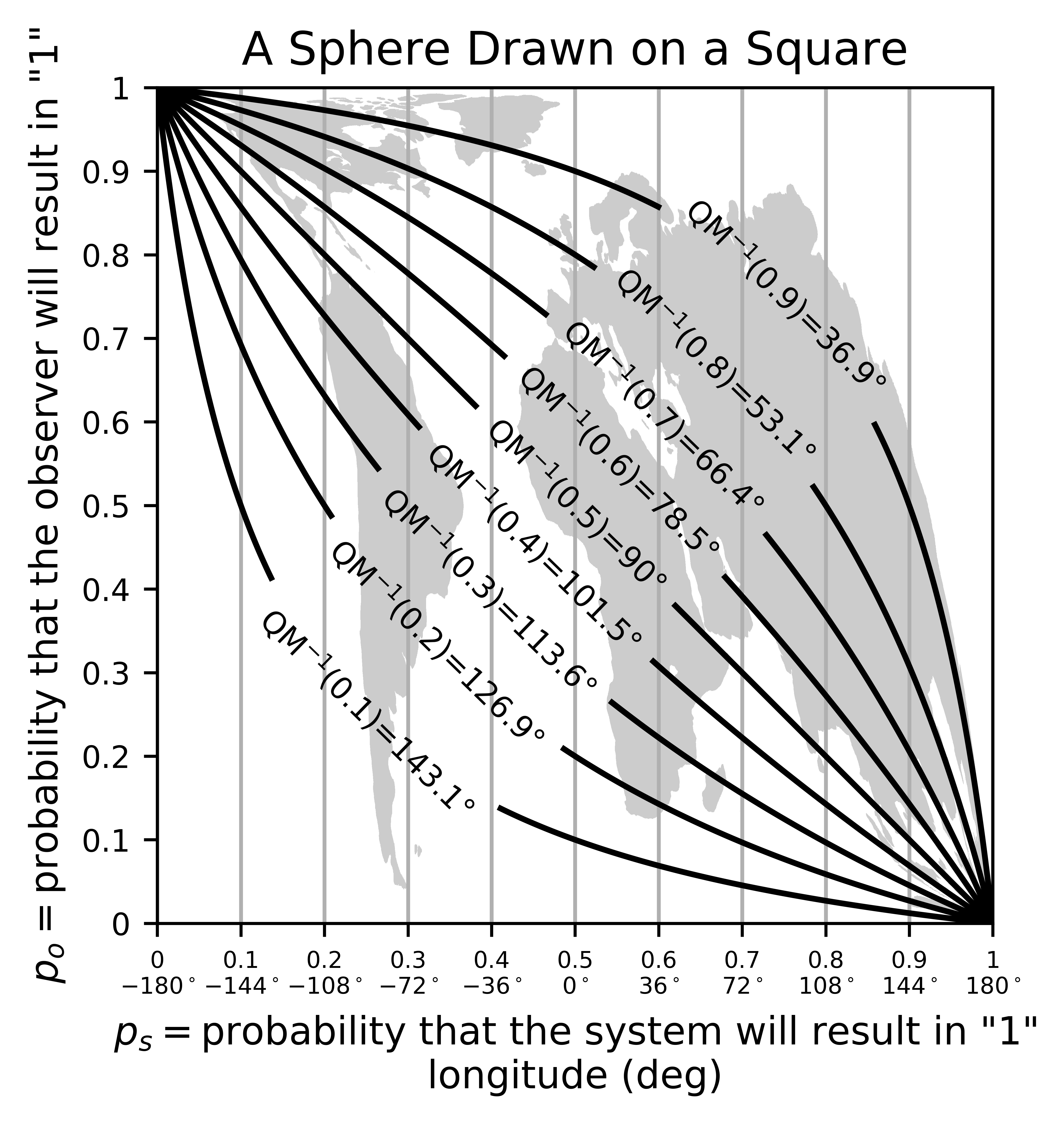}
\caption{Depicted is a graphical representation of the distance function described in section~\ref{distfunc}. 
The axes and contour lines are retained from Fig.~\ref{wootters}.  In order to find the distance between points on the unit square, we map to the corresponding points on the sphere and measure the angle subtended by these points and the center of the sphere.  The lines of latitude coincide with the contour lines of Fig.~\ref{wootters}.  The  mapping of a particular value of latitude to a particular contour line is chosen such that the resulting relationship between distance and probability matches the quantum mechanical probability function $\mbox{QM}(\theta)=\cos^2(\theta/2)$.}
\label{mercator}
\end{figure}

We define our distance function first by mapping the unit square to the sphere and then by employing the natural distance function on the sphere (i.e., the angle created by lines passing through the center of the sphere and the two points in question).  The particular form of the map we choose is given by
\begin{equation} 
\begin{split}
\theta(p_o,p_s) &=\mbox{QM}^{-1}\left[\mbox{CS}(p_s,p_o)\right]*(180^\circ/\pi)\\
\phi(p_o,p_s)&=(p_s-0.5)*360^\circ, \\
\end{split}
\label{transform}
\end{equation}
where $\mbox{QM}(\theta)\equiv\cos^2(\theta/2)$ is the quantum mechanical probability function, $\mbox{QM}^{-1}(p)=2\arccos\sqrt{p}$ is the inverse of $\mbox{QM}$, $\theta$ is the colatitude ($0^\circ$ at the north pole, $180^\circ$ at the south pole), and $\phi$ is the longitude.  This association of the unit square with points on the sphere is illustrated in Fig.\ref{mercator}.  The lines of latitude map to the contour lines of Fig.~\ref{wootters}.

We claim that an observer in this model will see a physics indistinguishable from quantum mechanics.  The argument is proof by construction.  The distance function given by Eqns.~(\ref{transform}) entails a state space (points on a sphere) with probability of projection of one state onto another given by $\mbox{QM}(\theta)$ where $\theta$ is the angle between preparation and measurement states.  This combination of state space and probability function is precisely the mathematics of a two-state quantum system.  Note that this argument works for other possible distance functions.  For example, Eqns.~(\ref{transform}) describe a linear relationship between $\phi$ and $p_s$; however, the connection to quantum mechanics is maintained for any continuous one-to-one map between $\phi$ and $p_s$.  


In the model we are considering, the observer manipulates degrees of freedom to choose preparation and measurement states by, for example, turning knobs on an experimental apparatus.  The calibration markings on these knobs are not provided to the observer;  rather, they are constructed using the results of the observer's measurements.  Thus, all choices of distance function compatible with quantum mechanics will be indistinguishable to the observer.  Therefore, we consider all such distance functions operationally equivalent for the purposes of this argument.

We have shown that our mechanism for passage is consistent with quantum mechanics.  We are now in a position to argue that quantum mechanics is a forced move given our mechanism for passage.  The final step relates to the status of statistical mixtures of preparation states.  It is a property of quantum mechanics that regardless of the complexity of a statistical source of quantum states, the source may be characterized by a finite description (i.e., the density matrix).  Were this not so, state determination of mixed states would be impossible (i.e., there would be no finite number of measurement settings sufficient to completely characterize the state).  We denote this property with the phrase ``mixtures have state'' and the abbreviation MHS.  

It may seem that MHS is a quirk of quantum mechanics.  We argue that in fact MHS is a requirement for any physics.  Earlier in the paper we argued that the concept of state preparation and measurement is implicit in the idea of physics.  Take it away, and we simply have a journal of unstructured interventions and observations.  Here we extend that argument to insist that the state description must be a finite function of the number of distinct output states, even in the case of statistical mixtures.  For an $N$-state system, this finite function is $N-1$ in the classical case, and it is $N^2-1$ in the quantum case. The argument is simply that physics could not proceed if such a finite function did not exist.  As mentioned above, state determination would be impossible.  In the absence of an operational method for characterizing states and making predictions about measurements, physics is impossible.  Thus, we argue that MHS is a necessary feature of any potential model of the world.

We now make the central conjecture of this paper:

\begin{conjecture}\label{walton}
Any choice of a distance function which satisfies MHS yields a physics indistinguishable from quantum mechanics.
\end{conjecture}

While we do not provide a proof of this conjecture here, we make a claim of plausibility by referring to Hardy's derivation of quantum mechanics from five reasonable axioms~\cite{hardy2001quantum}.  Hardy was able to show that if a physics has a state space such that the number of degrees of freedom of a mixed state is $N^2-1$ for an $N$-state system, then quantum mechanics is a forced move.  The question becomes then, why should a mixed state have $N^2-1$ degrees of freedom, as opposed to the more familiar $N-1$ of classical probability theory.  We claim that our argument for the necessity of a mechanism for time passage provides an answer to this question.  Since we need a mechanism for time flow, and our proposed mechanism entails $2(N-1)$ degrees of freedom for pure states, what remains is to adapt Hardy's proof to start with the dimensionality of pure states, rather than mixed states.

\begin{figure}
\includegraphics[width=0.5\textwidth]{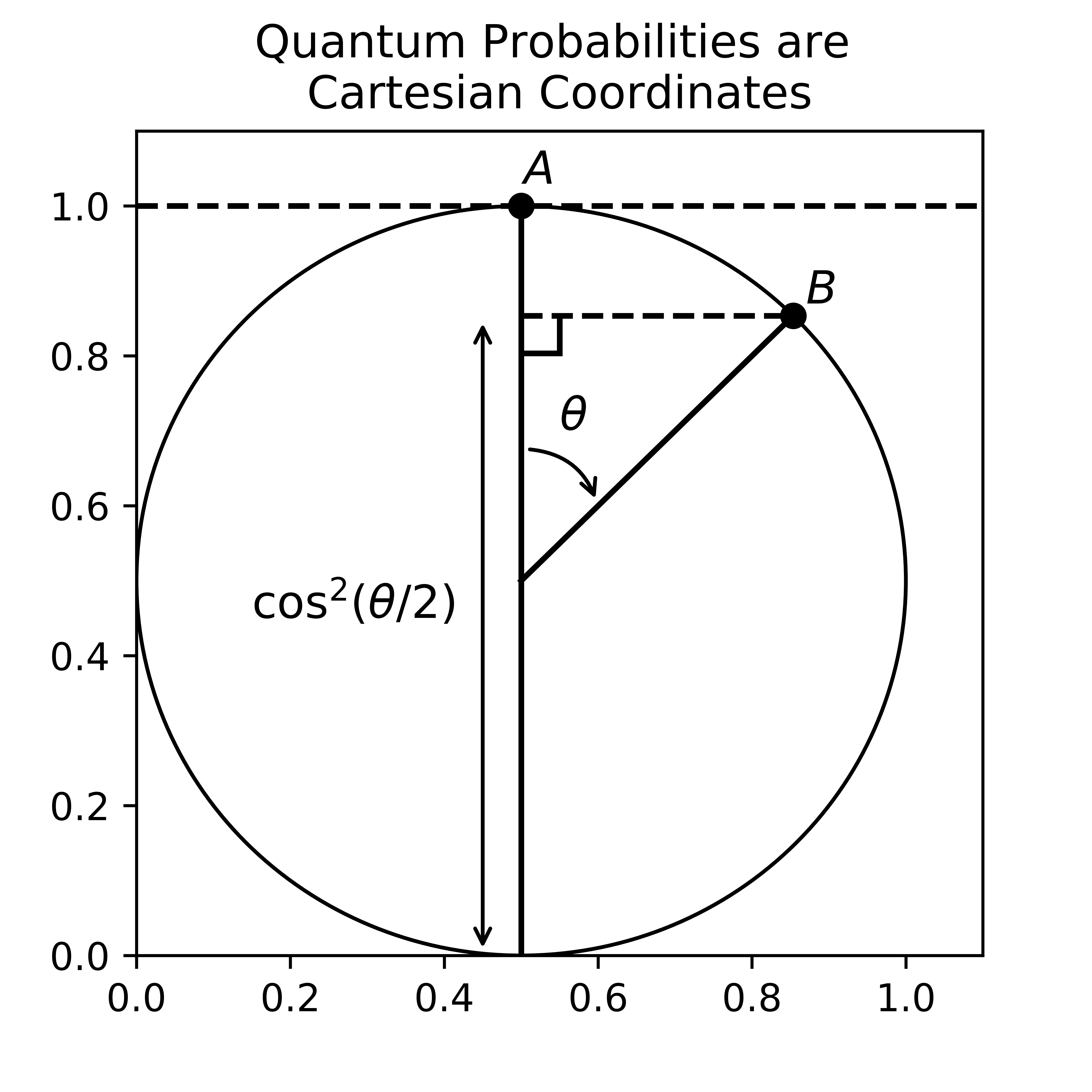}
\caption{
This figure demonstrates the connection between the quantum probability function $\cos^2(\theta/2)$ and the property that mixtures have state (MHS).  Given two pure qubits represented by points $A$ and $B$ on the Block sphere, the probability that $A$ will project onto $B$ is given by $\cos^2(\theta/2)$, where $\theta$ is the angle subtended by $A$, the center of the sphere, and $B$.  This probability is also the height of $B$ along the $y$-axis when the great circle defined by $A$ and $B$ is oriented such that $A$ is at the point $(0.5,1)$.  From this relation, it is clear that calculating a density matrix is equivalent to finding the center of mass of a system of point masses.  In each case, a potentially infinite description is condensed into a finite number of degrees of freedom.
}
\label{qmprob}
\end{figure}

Before concluding, we point out that the MHS constraint provides a particularly intuitive explanation of why the quantum probability function takes the form $\cos^2(\theta/2)$, as opposed to say, $\cos^4(\theta/2)$.  Figure~\ref{qmprob} shows a trigonometric identity which suggests a geometric interpretation of the quantum probability rule.  Consider two points on the Bloch sphere, $A$ and $B$.  If we wish to know the probability that the quantum state corresponding to $B$ will project onto the state corresponding to $A$, we may orient the great circle defined by $A$ and $B$ on the $x-y$ plane, such that the point $A$ is at the top with $y$ coordinate $1$.  The probability we seek is simply the $y$-value of the coordinates of $B$.  This association of probability with coordinate is true if and only if our probability function is $\cos^2(\theta/2)$.  Given this relationship, there is a direct analogy between finding the center of mass of an arbitrary number of point masses on the surface of a sphere and finding the density matrix corresponding to an arbitrary statistical ensemble of pure qubits.  We do not claim that this argument serves as a derivation of quantum mechanics; rather, it bolsters the plausibility of conjecture~\ref{walton}. 

\section{Conclusion}

In ref.~\cite{hardy2001quantum}, Hardy derives quantum mechanics from surprisingly innocuous axioms.  One way to summarize his accomplishment is to assert: If you have a reason to use more than N-1 degrees of freedom to represent an $N$-state system, then you need no longer puzzle about the form of quantum mechanics, as its precise structure is constrained by uncontroversial operational considerations.  This summary highlights the connection to our mechanism for time passage, since at the heart of the mechanism is correlated selection.  We arrived at correlated selection by looking for a way to encode the observer's identity in the relationship between before and after.  Crucially, this encoding requires that certain possible outcomes are removed from the set of all possible outcomes, such that the number of parameters needed to describe `after' is less than that needed to describe `before'.  It is an encouraging fact that the number of parameters specified by correlated selection (i.e., $2N-2$ for pure states) matches that of a pure quantum state.  Furthermore, this match motivates our conjecture that Hardy's derivation can be adapted to show that correlated selection makes quantum mechanics a forced move.

In the spirit of John Wheeler, we have argued that quantum mechanics arises from a distinction between system and observer.  To have an observer, we need naturalized time passage, described here as correlated selection in the context of a geometrized joint probability space.  To have a system, we require that mixtures have state.  If the hypothesized connection between our model of time passage and Hardy's work can be established, we are left with a novel approach to Wheeler's questions. ``How come existence?'' Because a distance function can be found that satisfies MHS. ``Why the quantum?'' Because all such distance functions are operationally equivalent to quantum mechanics.

\section{Future directions}

An obvious next step is to prove conjecture~\ref{walton}, either by adapting Hardy's proof or using some other method.  If the conjecture is wrong, and there are other distance functions which satisfy MHS, it would be interesting to investigate the properties of the resulting non-quantum mechanical model(s) of physics.  

We have proposed that quantum mechanics can be derived from probability theory and a conception of system and observer based on discrete mathematics. There is evidence that relativity (both special and general) can be derived in similar ways~\cite{Toffoli}.  Perhaps the long-sought unification of quantum mechanics and relativity can be achieved when we arrive at these theories from so simple a beginning.

\section{Acknowledgements}

Z.~W. was fortunate to learn of these issues in an undergraduate course taught by Simon Saunders.  Z.~W.~ also benefited from enlightening conversations with Chris Adami and Nicolas J.~Cerf.


\bibliography{bibl}

\begin{thebibliography}{10}%
\makeatletter
\providecommand \@ifxundefined [1]{%
 \@ifx{#1\undefined}
}%
\providecommand \@ifnum [1]{%
 \ifnum #1\expandafter \@firstoftwo
 \else \expandafter \@secondoftwo
 \fi
}%
\providecommand \@ifx [1]{%
 \ifx #1\expandafter \@firstoftwo
 \else \expandafter \@secondoftwo
 \fi
}%
\providecommand \natexlab [1]{#1}%
\providecommand \enquote  [1]{``#1''}%
\providecommand \bibnamefont  [1]{#1}%
\providecommand \bibfnamefont [1]{#1}%
\providecommand \citenamefont [1]{#1}%
\providecommand \href@noop [0]{\@secondoftwo}%
\providecommand \href [0]{\begingroup \@sanitize@url \@href}%
\providecommand \@href[1]{\@@startlink{#1}\@@href}%
\providecommand \@@href[1]{\endgroup#1\@@endlink}%
\providecommand \@sanitize@url [0]{\catcode `\\12\catcode `\$12\catcode
  `\&12\catcode `\#12\catcode `\^12\catcode `\_12\catcode `\%12\relax}%
\providecommand \@@startlink[1]{}%
\providecommand \@@endlink[0]{}%
\providecommand \url  [0]{\begingroup\@sanitize@url \@url }%
\providecommand \@url [1]{\endgroup\@href {#1}{\urlprefix }}%
\providecommand \urlprefix  [0]{URL }%
\providecommand \Eprint [0]{\href }%
\providecommand \doibase [0]{http://dx.doi.org/}%
\providecommand \selectlanguage [0]{\@gobble}%
\providecommand \bibinfo  [0]{\@secondoftwo}%
\providecommand \bibfield  [0]{\@secondoftwo}%
\providecommand \translation [1]{[#1]}%
\providecommand \BibitemOpen [0]{}%
\providecommand \bibitemStop [0]{}%
\providecommand \bibitemNoStop [0]{.\EOS\space}%
\providecommand \EOS [0]{\spacefactor3000\relax}%
\providecommand \BibitemShut  [1]{\csname bibitem#1\endcsname}%
\let\auto@bib@innerbib\@empty
\bibitem [{\citenamefont {Dennett}(1991)}]{dennett}%
  \BibitemOpen
  \bibfield  {author} {\bibinfo {author} {\bibfnamefont {Daniel~C.}\
  \bibnamefont {Dennett}},\ }\href@noop {} {\emph {\bibinfo {title}
  {Consciousness Explained}}}\ (\bibinfo  {publisher} {Little Brown \& Co.},\
  \bibinfo {year} {1991})\ p.\ \bibinfo {pages} {398}\BibitemShut {NoStop}%
\bibitem [{\citenamefont {Wheeler}(1990)}]{wheeler}%
  \BibitemOpen
  \bibfield  {author} {\bibinfo {author} {\bibfnamefont {John~A.}\ \bibnamefont
  {Wheeler}},\ }\bibfield  {title} {\enquote {\bibinfo {title} {Information,
  {P}hysics, {Q}uantum: {T}he {S}earch for {L}inks},}\ }in\ \href@noop {}
  {\emph {\bibinfo {booktitle} {Complexity, Entropy and the Physics of
  Information}}},\ \bibinfo {editor} {edited by\ \bibinfo {editor}
  {\bibfnamefont {Wojciech~H.}\ \bibnamefont {Zurek}}}\ (\bibinfo  {publisher}
  {Westview Press},\ \bibinfo {year} {1990})\BibitemShut {NoStop}%
\bibitem [{\citenamefont {Maudlin}(2007)}]{maudlin}%
  \BibitemOpen
  \bibfield  {author} {\bibinfo {author} {\bibfnamefont {Tim}\ \bibnamefont
  {Maudlin}},\ }\href@noop {} {\emph {\bibinfo {title} {The Metaphysics Within
  Physics}}}\ (\bibinfo  {publisher} {Oxford University Press},\ \bibinfo
  {year} {2007})\BibitemShut {NoStop}%
\bibitem [{\citenamefont {Saunders}(1998)}]{saunders}%
  \BibitemOpen
  \bibfield  {author} {\bibinfo {author} {\bibfnamefont {Simon}\ \bibnamefont
  {Saunders}},\ }\bibfield  {title} {\enquote {\bibinfo {title} {{Time, Quantum
  Mechanics, and Probability}},}\ }\href@noop {} {\bibfield  {journal}
  {\bibinfo  {journal} {Synthese}\ }\textbf {\bibinfo {volume} {114}},\
  \bibinfo {pages} {373--404} (\bibinfo {year} {1998})}\BibitemShut {NoStop}%
\bibitem [{\citenamefont {Mermin}(2013)}]{mermin}%
  \BibitemOpen
  \bibfield  {author} {\bibinfo {author} {\bibfnamefont {N.~David}\
  \bibnamefont {Mermin}},\ }\href@noop {} {\enquote {\bibinfo {title} {{QB}ism
  as {CB}ism: {S}olving the {P}roblem of ``the {N}ow''},}\ } (\bibinfo {year}
  {2013}),\ \Eprint {http://arxiv.org/abs/quant-ph/1312.7825}
  {arXiv:quant-ph/1312.7825 [quant-ph]} \BibitemShut {NoStop}%
\bibitem [{\citenamefont {Fuchs}\ \emph {et~al.}(2014)\citenamefont {Fuchs},
  \citenamefont {Mermin},\ and\ \citenamefont {Schack}}]{fuchs}%
  \BibitemOpen
  \bibfield  {author} {\bibinfo {author} {\bibfnamefont {Christopher~A.}\
  \bibnamefont {Fuchs}}, \bibinfo {author} {\bibfnamefont {N.~David}\
  \bibnamefont {Mermin}}, \ and\ \bibinfo {author} {\bibfnamefont
  {R{\"u}diger}\ \bibnamefont {Schack}},\ }\bibfield  {title} {\enquote
  {\bibinfo {title} {{A}n {I}ntroduction to {QB}ism with an {A}pplication to
  the {L}ocality of {Q}uantum {M}echanics},}\ }\href@noop {} {\bibfield
  {journal} {\bibinfo  {journal} {Am. J. Phys.}\ }\textbf {\bibinfo {volume}
  {82}},\ \bibinfo {pages} {749--754} (\bibinfo {year} {2014})},\ \Eprint
  {http://arxiv.org/abs/quant-ph/1311.5253} {arXiv:quant-ph/1311.5253
  [quant-ph]} \BibitemShut {NoStop}%
\bibitem [{\citenamefont {Wootters}(1996)}]{wootters}%
  \BibitemOpen
  \bibfield  {author} {\bibinfo {author} {\bibfnamefont {William~K.}\
  \bibnamefont {Wootters}},\ }\bibfield  {title} {\enquote {\bibinfo {title}
  {Is {T}ime {A}symmetry {L}ogically {P}rior to {Q}uantum {M}echanics?}}\ }in\
  \href@noop {} {\emph {\bibinfo {booktitle} {Physical Origins of Time
  Asymmetry}}},\ \bibinfo {editor} {edited by\ \bibinfo {editor} {\bibfnamefont
  {J.~J.}\ \bibnamefont {Halliwell}}, \bibinfo {editor} {\bibfnamefont
  {J.}~\bibnamefont {Perez-Mercader}}, \ and\ \bibinfo {editor} {\bibfnamefont
  {W.~H.}\ \bibnamefont {Zurek}}}\ (\bibinfo  {publisher} {Cambridge University
  Press},\ \bibinfo {year} {1996})\BibitemShut {NoStop}%
\bibitem [{\citenamefont {Wigner}(2001)}]{wigner}%
  \BibitemOpen
  \bibfield  {author} {\bibinfo {author} {\bibfnamefont {Eugene~P.}\
  \bibnamefont {Wigner}},\ }\bibfield  {title} {\enquote {\bibinfo {title}
  {Remarks on the {M}ind-{B}ody {Q}uestion},}\ }in\ \href@noop {} {\emph
  {\bibinfo {booktitle} {The Collected Works of Eugene Paul Wigner}}},\
  \bibinfo {editor} {edited by\ \bibinfo {editor} {\bibfnamefont {Jagdish}\
  \bibnamefont {Mehra}}}\ (\bibinfo  {publisher} {Springer},\ \bibinfo {year}
  {2001})\BibitemShut {NoStop}%
\bibitem [{\citenamefont {Hardy}(2001)}]{hardy2001quantum}%
  \BibitemOpen
  \bibfield  {author} {\bibinfo {author} {\bibfnamefont {Lucien}\ \bibnamefont
  {Hardy}},\ }\href@noop {} {\enquote {\bibinfo {title} {Quantum {T}heory
  {F}rom {F}ive {R}easonable {A}xioms},}\ } (\bibinfo {year} {2001}),\ \Eprint
  {http://arxiv.org/abs/quant-ph/0101012} {arXiv:quant-ph/0101012 [quant-ph]}
  \BibitemShut {NoStop}%
\bibitem [{\citenamefont {Toffoli}(1990)}]{Toffoli}%
  \BibitemOpen
  \bibfield  {author} {\bibinfo {author} {\bibfnamefont {Tommaso}\ \bibnamefont
  {Toffoli}},\ }\bibfield  {title} {\enquote {\bibinfo {title} {How {C}heap
  {C}an {M}echanics' {F}irst {P}rinciples {B}e?}}\ }in\ \href@noop {} {\emph
  {\bibinfo {booktitle} {Complexity, Entropy and the Physics of
  Information}}},\ \bibinfo {editor} {edited by\ \bibinfo {editor}
  {\bibfnamefont {Wojciech~H.}\ \bibnamefont {Zurek}}}\ (\bibinfo  {publisher}
  {Westview Press},\ \bibinfo {year} {1990})\BibitemShut {NoStop}%
\end{thebibliography}%

\end{document}